\documentclass[a4paper,fleqn]{cas-sc}

\usepackage[numbers]{natbib}

\def\tsc#1{\csdef{#1}{\textsc{\lowercase{#1}}\xspace}}
\tsc{AVB}
\tsc{AAZ}
\tsc{YMB}
\tsc{IVS}
\begin{document}
\let\WriteBookmarks\relax
\def\floatpagepagefraction{1}
\def\textpagefraction{.001}
\shorttitle{Comparative analysis of nonlinear elastic moduli of polystyrene, polycarbonate and PMMA}
\shortauthors{AV Belashov et~al.}
%\begin{frontmatter}

\title [mode = title]{Comparative analysis of nonlinear elastic moduli of polystyrene, polycarbonate and PMMA}                      
%\tnotemark[1,2]

%\tnotetext[1]{This document is the results of the research project funded by the Russian Science Foundation # 17-72-20201.}

%\tnotetext[2]{The second title footnote which is a longer text matter
%   to fill through the whole text width and overflow into
%   another line in the footnotes area of the first page.}

%[type=editor,
%auid=000,bioid=1,
%prefix=Sir,
%role=Researcher,
%orcid=0000-0001-7511-2910]
%\cormark[1]
%\ead{cvr_1@tug.org.in}
%\ead[url]{www.cvr.cc, cvr@sayahna.org}
%\credit{Conceptualization of this study, Methodology, Software}

\author[1]{A.V. Belashov}[orcid=0000-0002-0502-7769]
\author[1]{A.A. Zhikhoreva}
\author[1]{Y.M. Beltukov}[orcid=0000-0002-3648-0312]
\author[1]{I.V. Semenova}
\cormark[1]
\ead{irina.semenova@mail.ioffe.ru}

\address[1]{Ioffe Institute, 26, Polytekhnicheskaya, St.Petersburg, 194021, Russia}
\cortext[cor1]{Corresponding author}

%\cortext[cor2]{Principal corresponding author}
%\fntext[fn1]{This is the first author footnote. but is common to third
%  author as well.}
%\fntext[fn2]{Another author footnote, this is a very long footnote and
%  it should be a really long footnote. But this footnote is not yet
%  sufficiently long enough to make two lines of footnote text.}

%\nonumnote{This note has no numbers. In this work we demonstrate $a_b$
%  the formation Y\_1 of a new type of polariton on the interface
%  between a cuprous oxide slab and a polystyrene micro-sphere placed
%  on the slab.
%  }

\begin{abstract}
We present the comparative experimental analysis of  frequency dependencies of linear (Lam\'e) and nonlinear (Murnaghan) elastic moduli of polystyrene, PMMA and polycarbonate. The measurement methodology, based on the acousto-elastic effect, provided data on variations of these moduli in block samples of the polymers in the frequency range of 0.45--3 MHz. In all the three polymers the linear Lam\'e moduli demonstrated moderate rise with frequency, most pronounced rise was observed in modulus $\lambda$ of PMMA in about 35\%.  The frequency dependencies of Murnaghan moduli were considerably nonlinear.  
At higher frequencies above $\sim$1 MHz no significant variations of the Murnaghan moduli occurred, while at lower frequencies the absolute values of the moduli $l$ and $m$ demonstrated rapid rise, more pronounced for the modulus $l$. At the same time the absolute values of the modulus $n$ decreased and demonstrated a tendency to become positive at lower frequencies. Both linear and nonlinear moduli of PMMA had higher values than those of PC and PS, with the latter two demonstrating close values of both types of moduli. The potential origins of the differences in nonlinear elastic properties of the three polymers are discussed.
 
\end{abstract}

%\includegraphics[width=0.85\textwidth]{GraphicalAnstract3.png}
%\includegraphics{figs/grabs.pdf}

%\item shock wave can evolve into bulk strain soliton in nonlinear elastic waveguide
%\item soliton formation is due to elastic wave reflections from waveguide sidewalls
%\item waveguide viscosity provides efficient damping of high-oscillating wave components


\maketitle

\section{Introduction}

{\color{black}
Glassy polymers are among the most important engineering materials in use today. Characterized by their rigid, amorphous structure and operation below the glass transition temperature, these materials offer a compelling combination of high stiffness, dimensional stability, and ease of processing~\cite{Haward-physics-glassy-polymers-1997, Roth-polymer-glasses-2017}. These attributes have established them as reliable substitutes for metals across numerous applications, serving both as neat materials~\cite{OsswaldMaterial2012} and as robust matrices for composites~\cite{Mai-2006}. 
}

%Glassy polymers are among the most important engineering materials in use nowadays. They became a reliable substitution for metallic materials in a variety of industries, where they can be used both as neat materials and as matrices for composites.
%Products made of glassy polymers are widely applied in aerospace, automotive, gas and oil transportation, where they have to withstand high-power impacts. 

Notwithstanding the commercial importance of data on  material behavior and performance capability at such impacts, the fundamental understanding of physical/chemical origins of their behavior remains of high demand. The design of products made of glassy polymers requires detailed information on how these materials respond when being deformed. 
The response of glassy polymers to mechanical loads is strongly dependent on loading rate, temperature and thermomechanical prehistory of the material \cite{Spathis2001,OsswaldMaterial2012,Siviour2016}. These materials often demonstrate nonlinear viscoelastic properties, which provoke their complex mechanical behavior, and knowledge of these properties is crucial for reliable operation of products \cite{Spathis2001}. 
The state-of-the-art modeling of the deformation behavior of glassy polymers requires accounting for their viscoelastic-viscoplastic nature, which is characterized by a strong dependence of mechanical properties on temperature, strain rate, and hydrostatic pressure \cite{Aoyagi2025}. 
Besides that, the mechanical behavior of polymer materials may be governed by changes in molecular orientation in the course of deformation. Such features as orientation of polymer chains, free volume, presence of defects or fillers can affect the viscoelastic properties. 
The molecular relaxation processes were suggested to be related to
viscoelastic properties of glassy polymers, which can be characterized by frequency and temperature dependence of the elastic moduli \cite{Coda_wave}.

Polystyrene (PS), poly(methyl methacrylate) (PMMA), and polycarbonate (PC) are most widely used glassy amorphous polymers that exhibit distinct nonlinear elastic and viscoelastic properties, especially under varying frequencies, temperatures, and strain rates. At low strains these polymers are often treated as linearly elastic, however their performance is dominated by nonlinearities arising from their molecular structures and chain dynamics. The mechanical response of these polymers differs significantly due to their side-chain configurations and chain flexibility \cite{Lebedev1978}. At the same time these polymers were shown to exhibit similar intrinsic behavior in compression \cite{Melick2003}.

The linear elastic properties of these polymers have been studied sufficiently well. The strain-rate dependence of the Young's modulus was analyzed both theoretically and experimentally \cite{PMMA-PC-rate2001,PMMA-PS-rate2006,Mulliken2006} and its
noticeable increase with strain rate varied in a wide range has been observed in all the three polymers \cite{PMMA-PC-rate2001,PMMA-PS-rate2006}.  
The frequency and temperature dependence of linear elastic moduli have been examined in a number of studies. The frequency dependence of the elastic moduli of these polymers has been demonstrated  
\cite{Capodagli-2008-isothermal-viscoelastic-properties,Yadav-2020-effect-thermomechanical-couplings,Lagakos1986} and was suggested to be related to molecular relaxation processes  \cite{Lagakos1986}. 
The observed variations in the Young's modulus below glass transition temperature with frequency were typically not exceeding 100 \%, with general tendency of rise with frequency.

Much less attention was paid so far to nonlinear viscoelastic properties of glassy polymers. However, account for these properties can be particularly important for these materials, since they may experience  relatively large deformations under  small stresses, in contrast to traditional crystalline solids like metals.
%
%While the small-strain elasticity of glassy polymers is well characterized, their nonlinear finite-strain viscoelasticity remains comparatively underexplored. This regime is particularly consequential in glassy systems, where stress-activated molecular mobility can lead to time-dependent viscoelastic deformation far below the macroscopic yield point—a complexity generally absent in traditional crystalline solids like metals.
%
Some models have been proposed to describe the nonlinear viscoelasticity of glassy polymers with more or less satisfactory predictive capabilities, see e.g. \cite{Spathis2001,Aoyagi2025,Adolf2004}. The major problem in validation of these models is the deficiency of experimental data on nonlinear elastic properties of the specific materials and their dependence on various factors, such as wave frequency, temperature, etc. Among other aspects the lack of a routine methodology providing data on these properties limits progress in this direction.  

The Murnaghan's model \cite{Murnaghan}, which describes the nonlinear elastic behavior of an isotropic solid material, is currently utilized most often. In this model the nonlinear elastic behavior of a material is described using a set of five parameters: two linear elastic (Lam\'e) moduli $\lambda$ and $\mu$, and three nonlinear ones $l, m, n$. 
The elastic energy density is expressed as:
\begin{equation}
	\Pi = \frac{\lambda+2\mu}{2}I_1^2 - 2\mu I_2 + \frac{l + 2m}{3}I_1^3 - 2m I_1 I_2 + n I_3,
	\label{eq:pot_en}
\end{equation}
where
 $I_1 = \operatorname{tr}\mathbf{E}$, $I_2 = [ \left(\operatorname{tr}\mathbf{E}\right)^2 - \operatorname{tr}(\mathbf{E}^2)]/2$, and $I_3 = \det \mathbf{E}$ represent the invariants of the Green–Lagrange strain tensor $\mathbf{E}$.
Note that the parameters $l, m, n$ do not have a distinct physical meaning and were introduced in the model to describe deviation of the stress-strain relationship from the linear behavior. However, allowance for these parameters becomes crucial for correct description and prediction of a broad range of phenomena, in particular the material response to high dynamic loads. An account for nonlinear elastic properties of polymer materials is necessary for correct analysis of evolution of elastic waves, their interaction,  and estimation of variations of wave velocity with the applied stress or temperature \cite{Epoxy-moduli1990}, as well as for prediction of generation of nonlinear strain solitary waves \cite{Samsonov2001,WaMot2019,belashov2018indirect,WaMot2022}.   

The data on nonlinear elastic parameters of polystyrene, polycarbonate and PMMA have been reported elsewhere, see, \cite{Hughes-Kelly1953,Nlin-moduli-PC-PMMA1978,Nlin-moduli-PS-PMMA1989,Moduli_polycarb,PMMA-nonlin2016}. However, the dependencies of the moduli on frequency and temperature has been revealed only recently. Our recent research on frequency and temperature dependencies of Murnaghan moduli $l,m,n$ in polystyrene \cite{IJNLM2024,Polymers2025} demonstrated significant rise of the absolute values of the moduli $l$ and $m$ with decreasing frequency below about 1 MHz, as well as an increase with temperature. 

\textcolor{black}{In this paper we present a comparative analysis of the frequency dependencies of the $m,n$ moduli in the range of 0.4--3 MHz and $l$ in the range of 0.5--3 MHz in the three commonly used glassy polymers, polystyrene, polycarbonate and PMMA.} The potential origins of the observed differences in nonlinear elastic properties of the three polymers are discussed.

\section{Experimental procedure}

\subsection{Materials and measurement methodology}

Experiments were performed on block samples of commercial PMMA, polycarbonate (PC) and styrene copolymer with ethylene glycol dimethacrylate (EGDMA) at the concentration of 10\% (hereinafter referred to as polystyrene, PS). The latter material was manufactured by the Dzerzhinsk Enterprise of Organic Synthesis. Samples were fabricated in the form of bars, 50 mm long, 10×10 mm$^2$ in cross section.

Note that due to significant attenuation of shear waves in polycarbonate, for measurements with this kind of waves the PC sample length was reduced from 50 to 10 mm. That allowed us to roughly estimate the linear and nonlinear elastic moduli of polycarbonate, although the decrease of wave propagation distance reduced the measurement accuracy.

%The investigations used block samples of commercial styrene copolymer with ethylene glycol dimethacrylate (EGDMA) at a 10\% concentration that were made by the Dzerzhinsk Enterprise of Organic Synthesis. We'll call it polystyrene from now on.
%PC AND PMMA

\begin{figure}[]
	\centering
	\includegraphics[width=10cm]{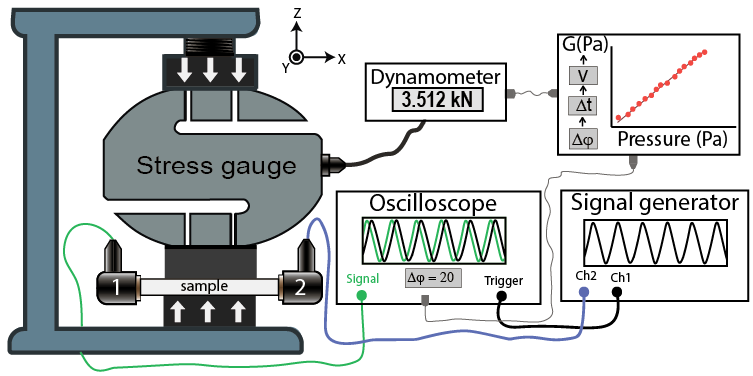}
	\caption{Experimental setup used for measurements of nonlinear elastic moduli of polymer samples.}
	\label{FigSetup2}
\end{figure}

The methodology used to measure the nonlinear elastic (Murnaghan) moduli of the samples was
based on the acousto-elastic effect and utilized an analysis of variations in velocities of longitudinal and shear ultrasonic waves in the sample with the applied static stress \cite{Hughes-Kelly1953,PoTe2021}. 
The experimental setup (shown schematically in Fig.~\ref{FigSetup2}) comprised a jaw vice, a high-precision stress gauge, piezoelectric transducers, a pulse generator and an oscilloscope. 
A sample was clamped in the jaw vice and was subjected to the uniaxial static pressure, which was gradually increased manually in the range of 0–12~MPa under control by the stress gauge. The applied pressure in the chosen range did not cause any irreversible deformations or significant changes in mechanical properties of the samples  in the course of measurements. 

The same experimental setup was used also for measurements of the linear elastic (Lam\'e) moduli by evaluation of velocities of longitudinal and shear ultrasonic  waves under atmospheric pressure. These moduli were calculated from data on the transition time of the wave packet of a specific frequency  through the sample. 

Shear ultrasonic waves were generated and detected using V154-RB piezoelectric transducers (Olympus, USA). LS814M-800 transducers (Amati Acoustics, Russia) were used for longitudinal waves in the sub-megahertz range, and P121 transducers (Amati Acoustics, Russia) were used for measurements at frequencies above 1 MHz. Wave packets of the desired shape and carrier frequency or sinusoidal waves were generated using an AM300 Dual Arbitrary Generator (Rohde\&Schwarz), the detection was performed using an RTB2002 oscilloscope (Rohde\&Schwarz). In all experiments, signal averaging over 16 measurements was performed to improve measurement accuracy.

%\begin{figure}[]
%	\centering
	%\includegraphics[width=12cm]{Setup5.png}
%	\caption{Experimental setup used for measurements of nonlinear elastic moduli of polymer samples.}
%	\label{FigSetup}
%\end{figure}

\subsection{Measurements of linear elastic moduli}

%Linear and nonlinear elastic moduli of polymers were measured using PS, PC, and PMMA samples shaped as $\approx$10$\times$10$\times$50 mm bars. 
For measurements of the linear elastic moduli we generated wave packets of  longitudinal or shear ultrasonic waves at a given carrier frequency and a Gaussian envelope. The width of the Gaussian envelope equaled to 8 \textmu s, which corresponded to a spectral range of approximately 65~kHz. That enabled measurements of longitudinal and shear wave velocities from 400~kHz to 3~MHz with a spectral measurement step starting from 125~kHz at low frequencies. Measurements of the time required for the wave packet to pass through the sample allowed for calculation of wave velocity and its dependence on frequency.

\subsection{Measurements of noninear elastic moduli}

%To measure Murnaghan's nonlinear elastic moduli, an experimental setup was used to estimate them based on the acoustoelastic effect by measuring the velocity of longitudinal and shear ultrasonic waves when static pressure is applied to the sample in a direction perpendicular to the direction of ultrasonic wave propagation. \cite{Hughes-Kelly1953,PoTe2021} 
The absolute values of longitudinal and shear wave velocities in the three materials at atmospheric pressure were obtained while measuring the linear elastic moduli. Therefore, to obtain the dependencies of wave velocities on the applied static pressure ($P$) we performed estimations of variations of the time spent by the wave for propagation through the sample ($\Delta t_P$). To achieve this and to ensure frequency-dependent measurements, in these experiments we used continuous sinusoidal waves at given frequencies in the range of 0.4–3 MHz. Generation of a sinusoidal wave at the input edge of the sample and its detection at the output edge allowed estimating the phase $\phi$ of the transmitted signal. Assuming that measurement was performed at a known single frequency $f$ of the ultrasonic wave, the phase change of the signal recorded at a static pressure ($P$) $\Delta \phi_P = \phi_P - \phi_0$  allows calculating the corresponding change in the wave propagation time: $\Delta t_P = \Delta \phi_P / 2 \pi f$. Taking into account the known length of the sample $L\approx$ 50 mm and the time of wave propagation at zero pressure $t_0 = L / V_0$, the velocity of ultrasonic wave of the given frequency at the static pressure $P$ can be calculated as $V_P = L / (t_0 + \Delta t_P)$.
Therefore, the dependence of wave velocity on static pressure $V_P$, can be obtained from measurements of the phase shift of the continuous sinusoidal wave upon increase of static pressure applied to the sample: $V_P = L / (t_0 + \Delta \phi_P / 2 \pi f)$.

%Thus, taking into account the previously measured propagation velocities of ultrasonic waves in the absence of static pressure on the sample $V_0$, the measurement of the dependence of these velocities on static pressure $V_P$, can be reduced to measurement of the change in the phase of a continuous sinusoidal wave upon increase of ststic pressure applied to the sample:  .%at one end of the sample when it is generated at the other end. 
Unlike measurements using a Gaussian wave packet, where dispersion and pressure can significantly alter its shape during propagation through the sample, measurements utilizing single-frequency sinsodial waves allow for both precise determination of the ultrasonic signal frequency and accurate evaluation of changes in the wave velocity at increasing static pressure.

\begin{figure}[]
	\centering
	\includegraphics[width=14.5cm]{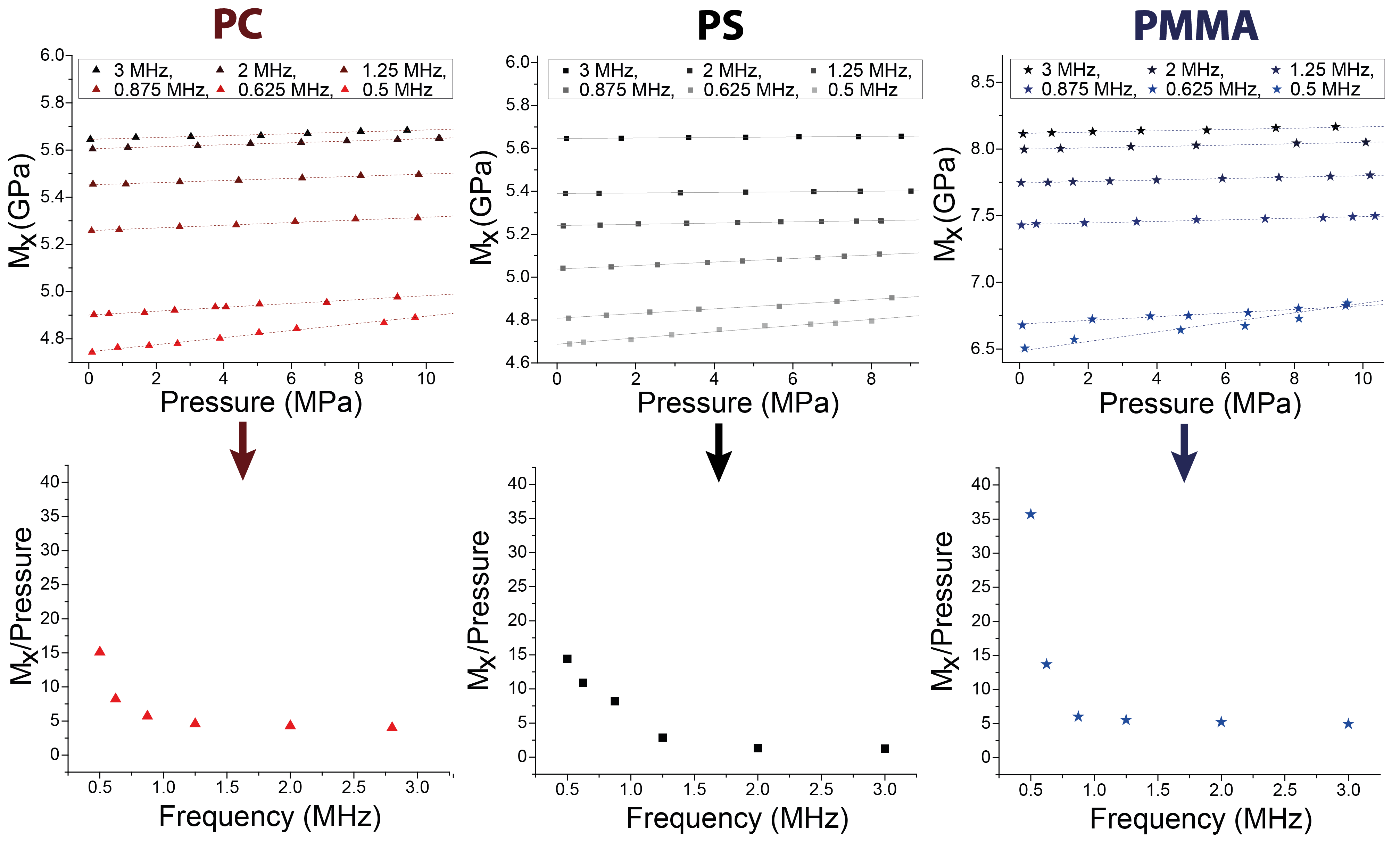}
	\caption{Flowchart of experimental data processing for the three polymer samples. First row: effective elastic modulus $M_x$ as a function of applied pressure, second row: dimensionless slope coefficient $\alpha_x$ as a function of wave frequency.}
	\label{calculation}
\end{figure}

To determine the three nonlinear Murnaghan elastic moduli, the static pressure was applied to the sample along the $Z$ axis (see Fig.~\ref{FigSetup2}) and was increased gradually after each measurement. Three measurement cycles were performed for each sample. Each cycle comprised measurements of velocities of one type of ultrasonic waves (longitudinal waves ($V_x$) and shear waves with an oscillation vector directed perpendicular ($V_y$) and parallel ($V_z$) to the direction of static pressure $Z$) as a function of applied pressure. The three cycles were repeated for each frequency of the ultrasonic waves in the range of 0.4–3 MHz.

The data obtained  were used to calculate the dependence of the effective elastic moduli $M_i(P)=\rho V_i(P)^2$. In the restriction of the expansion of the elastic energy density in terms of invariants of the Green–Lagrange strain tensor to the third-order terms these moduli are supposed to depend linearly on the static pressure applied to the sample $M_i(P) = M_i(0) + \alpha_i P$. In the absence of static pressure the effective elastic moduli $M_i(0)$ were determined from the linear elastic moduli as $M_x(0)=\lambda+2\mu=\rho V_p^2$ and $M_y(0)=M_z(0)=\mu=\rho V_s^2$. The calculated values of the three dimensionless slope coefficients $\alpha_i$ along with the obtained values of the linear elastic moduli $\lambda$ and $\mu$ were used to calculate the nonlinear Murnaghan moduli $l$, $m$ and $n$ using the system of equations \cite{PoTe2021}:
\begin{gather}
    l = -\frac{3 \lambda +2 \mu }{2} \alpha_x-\frac{\lambda  (\lambda +\mu ) }{\mu}(1+2 \alpha_y)+\frac{\lambda^2 }{2 \mu}(1-2 \alpha_z),  \label{eq:l}\\
    m = -2 (\lambda +\mu ) \left(1+\alpha_y\right)+\lambda  \left(1-\alpha_z\right),  \label{eq:m}\\
    n = -4 \mu \left(1+\alpha_y-\alpha_z\right),  \label{eq:n}
\end{gather}
Mention that the above system of equations takes into account the specimen elongation under static pressure by a factor of $L \nu P / E$, where $L$ is the specimen length, $\nu$ is Poisson's ratio, and $E$ is Young's modulus. 
%Thus, the scheme of experimental procedure for determining nonlinear elastic moduli is shown in Figure \ref{calculation}.

%The calculation of the whole set of nonlinear elastic moduli was performed for each target frequency 0.45-3 MHZ and for each studied materials: PC, PS, PMMA. 

Figure \ref{calculation} presents an illustration of the data processing flowchart on the example of experimental data on pressure- and frequency-dependent variations of longitudinal wave velocities in  samples of the three polymers. 
%The subsequent evaluation of $M_x(P)$ dependences and estimation of dimentionless slopes $\alpha_x$ as a function of ultrasonic wave frequency for three studied materials are also shown in figure \ref{calculation}. 
The data in Fig.~\ref{calculation} clearly demonstrates that the decrease in frequency $f$ resulted in more prominent dependence of longitudinal wave velocity on applied pressure featuring inverse dependence of the slope coefficient $\alpha_x$ on the ultrasonic wave frequency $f$.

\section{Results and discussion}

\subsection{Analysis of the frequency dependence of linear elastic moduli}

\begin{figure}[]
	\centering
	\includegraphics[width=13.5cm]{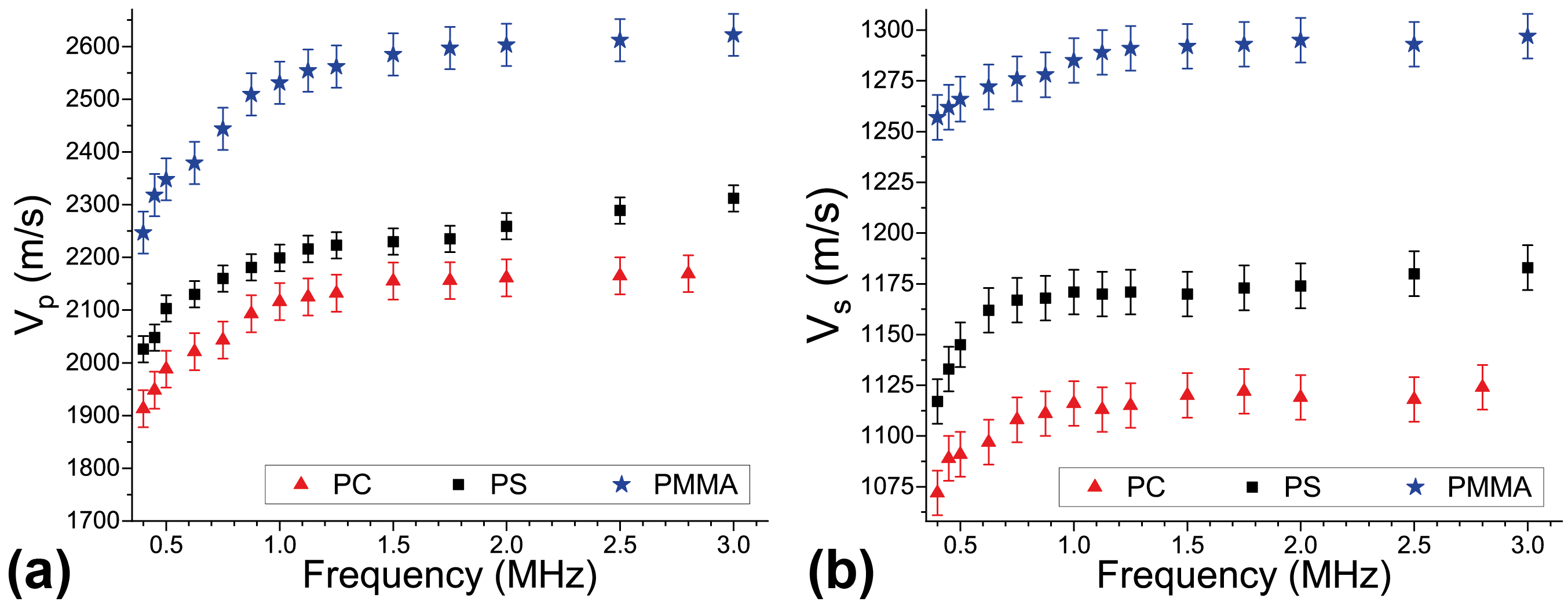}
	\caption{Frequency dependencies of (a) longitudinal and (b) shear wave velocities in samples of PC, PS, and PMMA in the range of 0.45-3 MHz.}
	\label{velocities}
\end{figure}

\begin{figure}[]
	\centering
	\includegraphics[width=13.5cm]{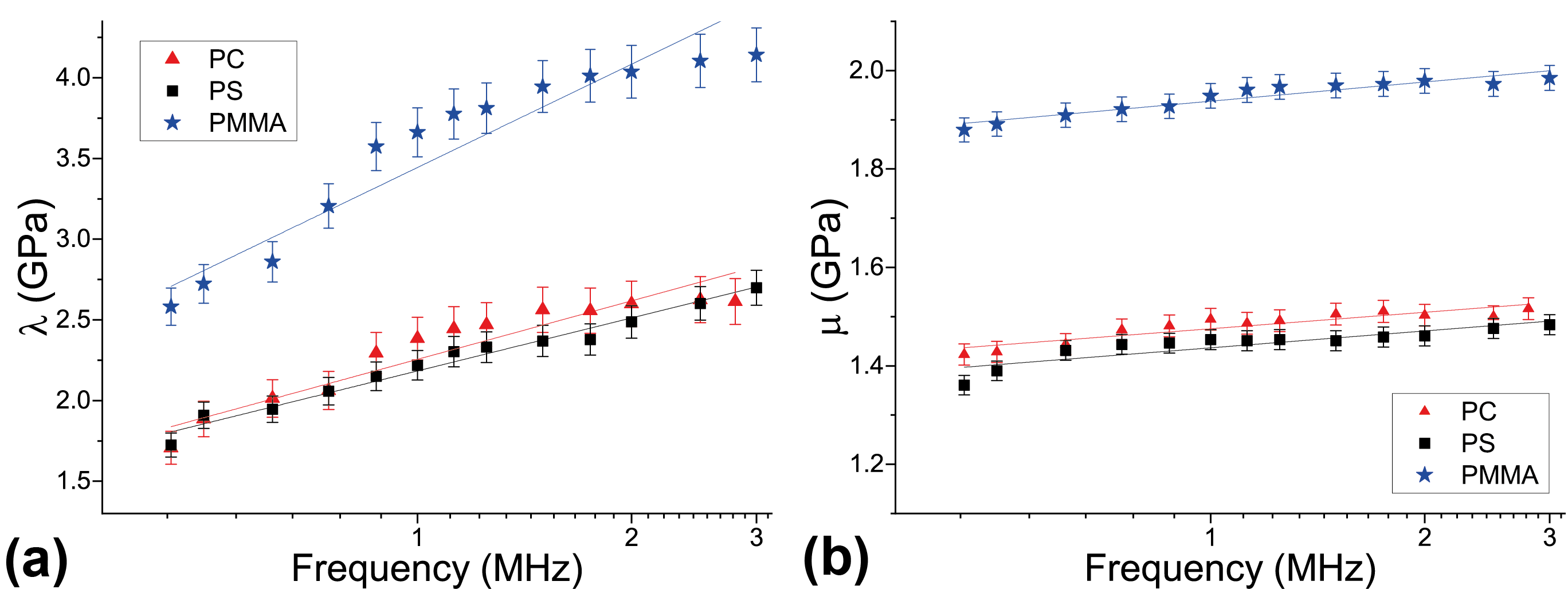}
	\caption{Linear elastic moduli (a) $\lambda$ and (b) $\mu$ of PC, PS, and PMMA in the frequency range of 0.45-3 MHz.}
	\label{linearmoduli}
\end{figure}

Figure \ref{velocities} presents data on longitudinal and shear wave velocities in PC, PS and PMMA obtained at atmospheric pressure as a function of wave frequency. The corresponding frequency dependencies of the calculated linear Lame elastic moduli are shown in Fig.~\ref{linearmoduli}.

The results obtained demonstrate significant differences in the velocities and elastic moduli $\lambda$ and $\mu$ between PMMA and the two other polymers, PC and PS. Although all the three materials are amorphous thermoplastics, differences in their molecular structure directly affect their mechanical properties and dictate how mechanical vibrations propagate through the samples.

\begin{figure}[]
	\centering
	\includegraphics[width=13cm]{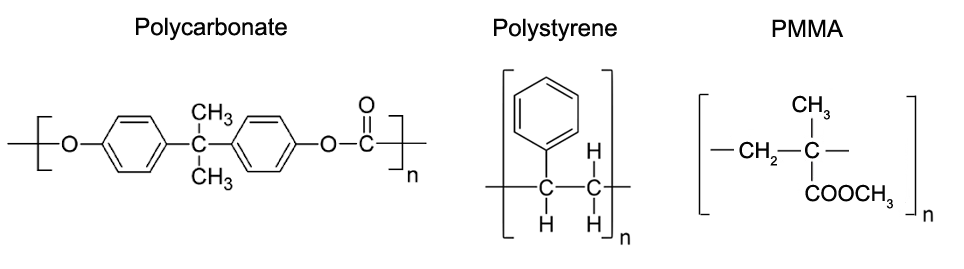}
	\caption{Molecular structures of PC, PS and PMMA.}
	\label{molstructures}
\end{figure}

\textcolor{black}{Molecular structure profoundly dictates the mechanical properties of polymers, with key factors including molecular weight, branching, cross-linking, and crystallinity \cite{Relation2017}. The molecular structures of PMMA, PS and PC differ significantly in their backbones and functional groups, as illustrated in Fig.~\ref{molstructures}. 
PMMA has a methyl group (–CH$_3$) and a methyl ester group (–COOCH$_3$) attached to the same carbon atom in the chain.
These bulky side groups prevent the chains from packing tightly into crystals, making it amorphous, but  somewhat brittle. The presence of a methyl group (–CH$_3$) within the polymer structure prevents tight packing in a crystalline manner and restricts free rotation around the C-C bonds \cite{PMMA2024}.
In PS a large, hexagonal phenyl group (benzene ring) is attached to every other carbon atom on the backbone. These massive structures act like ``anchors'', making the polymer rigid and glassy.
Unlike PS and PMMA, which have a simple carbon-only backbone, PC incorporates carbonate groups (–O–(C=O)–O–) and aromatic rings directly into the main chain.
The aromatic rings in the backbone provide high stiffness, while the carbonate linkages allow for some molecular flexibility. This combination enables PC to absorb a high-velocity impact without shattering, under which PMMA or PS would crack.
PC is 10–30 times tougher and more impact-resistant than PMMA and PS, while PMMA is more rigid, but more brittle than PC. PS is generally fragile and more brittle compared to both PC and PMMA. The cross-linkage by a linking agent can enhance its rigidity preventing chain slipping. 
}

The greater stiffness of PMMA can be attributed to several factors, such as a denser packing compared to PS and significant dipole-dipole interactions due to the presence of polar ester side groups (–COOCH$_3$), which lead to the formation of a more rigid network compared to polystyrene, where presence of nonpolar aromatic rings enable only weak intermolecular van der Waals forces. The massive aromatic rings in the backbone of polycarbonate and more flexible bonds between them facilitate slower propagation of elastic waves and significant attenuation of shear waves.

%Since the determination of nonlinear elastic moduli use the data on the linear elastic moduli and ultrasonic waves velocity in the absence of pressure, measurements of these parameters for PS, PC, and PMMA samples were performed over the entire frequency range of interest (Fig.~\ref{velocities} and \ref{linearmoduli}).

\subsection{Analysis of the frequency dependence of nonlinear elastic moduli}

\begin{figure}[]
	\centering
	\includegraphics[width=14cm]{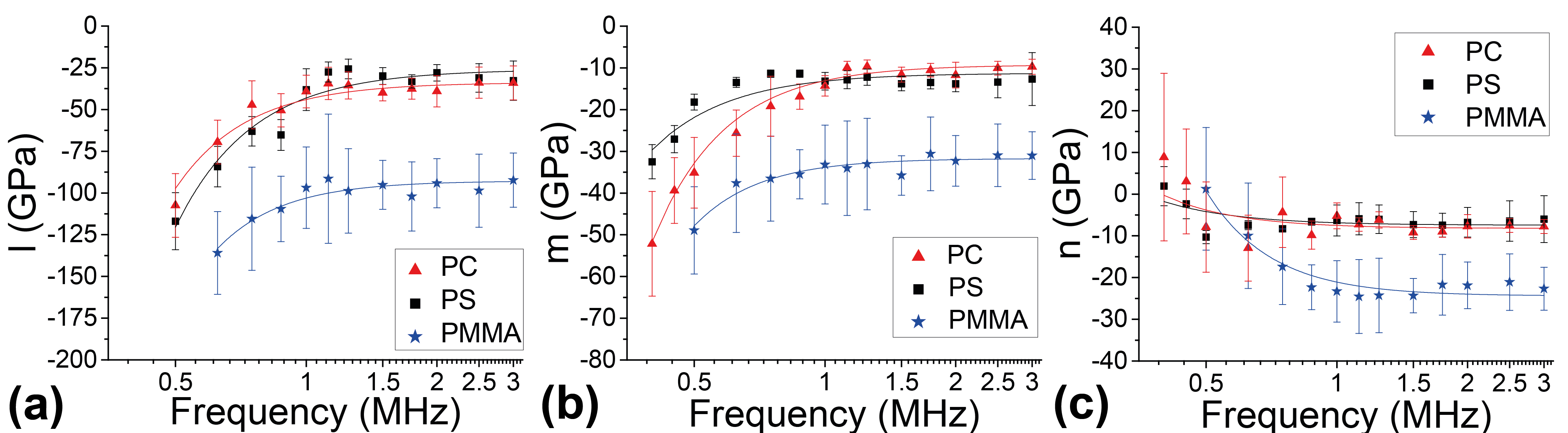}
	\caption{Fitting of the frequency dependencies of  nonlinear elastic moduli  $l$ (a),  $m$ (b) and  $n$ (c) under assumption of a single dominating relaxation process.}
	\label{nonlinearmoduli}
\end{figure}

%Using this measurement technique to determine the frequency dependences of linear and nonlinear elastic moduli in the frequency range from 0.45 to 3 MHz, measurements were performed for three types of polymers: polystyrene, polycarbonate, and PMMA. 
The obtained frequency dependencies of nonlinear elastic moduli $l$, $m$, $n$ of PC, PS and PMMA are presented in Fig.~\ref{nonlinearmoduli}.  
As can be seen from the data obtained, for all the three  polymers, variations of the nonlinear moduli are significantly more profound than those of the linear moduli (see Fig.~\ref{linearmoduli}). The average variations in the linear elastic moduli of these materials with frequency from 0.5 to 3 MHz amounted to $\sim$34\% for $\lambda$ modulus and $\sim$6\% for $\mu$ modulus. Meanwhile the nonlinear elastic moduli in this frequency range varied by a factor of 1.5–2.5 (see Table \ref{tablevariation}). The most profound changes in both absolute and relative values were observed in the nonlinear modulus $l$; the polymer demonstrating the greatest variations in nonlinear elastic moduli was polycarbonate. It is also worth noting that among the three polymer materials, significant differences between PMMA and the other two polymers were observed both in linear, by 30–50\%, and in nonlinear elastic moduli, where differences reached 2–3 times. 

As we mentioned above it can be associated with dipole interaction due to the presence of ester side groups –COOCH$_3$, which apparently impact not only the linear elastic properties but the nonlinear ones also. Moreover, the strong frequency dependence of nonlinear elastic moduli was observed for PMMA samples as well.

\begin{table}[]
\centering
\caption{Relative variations of linear and nonlinear elastic moduli of PC, PS and PMMA in the frequency range of 0.5-3 MHz.} 
\begin{tabular}{l ccccc}
\hline
\begin{tabular}[c]{@{}l@{}}Variations of linear \\ and nonlinear \\ elastic moduli:\end{tabular} & \multicolumn{1}{l}{$\delta \lambda$ (\%)} & \multicolumn{1}{l}{$\delta \mu$ (\%)} & \multicolumn{1}{l}{$\delta l$ (\%)} & \multicolumn{1}{l}{$\delta m$ (\%)} & \multicolumn{1}{l}{$\delta n$ (\%)} \\ \hline
PC                                                                             & 31.2                                       & 5.9                                    & 155.3                                & 165.5                                & 108.6                                \\ 
PS                                                                             & 34.5                                       & 6.5                                    & 190.2                                 & 51.2                                 & 62.2                                \\ 
PMMA                                                                           & 38.8                                       & 4.8                                    & 153.1                                & 64                                   & 152.8                                \\ \hline
\end{tabular}
\label{tablevariation}
\end{table}

% \textcolor{red}{The obtained frequency dependencies can be further analysed assuming the relaxation model similar to a generalized Maxwell model~\cite{Tschoegl-1989-phenomenological-theory-linear} which describes frequency dependence of the linear elastic moduli \cite{Garbuzov-2024}. In the case of a single dominating relaxation process the frequency dependence of the measured dynamic nonlinear moduli can be described as:}

The experimentally determined frequency dependence of the nonlinear elastic moduli exhibits a pronounced trend toward increasingly negative values with decreasing frequency. The dominant factor governing this frequency dependence is the presence of relaxation processes and their sensitivity to the applied pressure~\cite{Garbuzov-2024}. The observed behavior indicates that these relaxation processes occur in the sub-MHz range. 

For the glassy polymers under investigation, one should anticipate not a single relaxation mechanism, but rather a broad spectrum of relaxation times accompanied by complex mutual interactions among the corresponding processes. Consequently, for a more detailed analysis of the measured dependence, we employ the following empirical power-law dependence:
\begin{align}
    l(\omega) &= l_\infty + l_1\,\omega^{-\gamma},   \label{eq:lw} \\
    m(\omega) &= m_\infty + m_1\,\omega^{-\gamma},   \label{eq:mw}  \\
    n(\omega) &= n_\infty + n_1\,\omega^{-\gamma},   \label{eq:nw} 
\end{align}
where $l_\infty$, $m_\infty$, and $n_\infty$ denote the high-frequency limits of the corresponding nonlinear moduli, and $l_1$, $m_1$, and $n_1$ represent amplitudes of the frequency-dependent contributions characterized by the exponent $\gamma$.

\color{black}
Before applying the fitting procedure, it should be noted that the lowest frequency range of the data may be subject to distortion arising from several mechanisms.
\color{black}
%Note that low-frequency data can be distorted due to several factors. 
One of them is the occurrence of multiple reflections of ultrasonic waves in the course of their propagation through the sample, which happens when the attenuation length is greater than or roughly equal to twice the sample thickness. Another factor is the plane-wave dispersion in the waveguide, which becomes significant when the wavelength approaches the size of the waveguide cross-section. To avoid influence of these factors on the data obtained, we have set a low-frequency cutoff at 0.5~MHz for the modulus $l$ and at 0.4~MHz for the moduli $m$ and $n$, thereby defining the frequency range in which the measurements were trustworthy.

A global fit of the experimental data enabled the determination of $\gamma$ for each polymer. The resulting values are summarized in Table \ref{tablefitting}.
\color{black}
For all studied polymers, the frequency-dependent nonlinear coefficient $l_1$ has a higher absolute value than $m_1$ and $n_1$. It corresponds to the fact that the longitudinal sound velocity is the most sensitive to applied pressure, and the coefficient $\alpha_x$ influences only the nonlinear modulus $l$, as stated in Eqs.~(\ref{eq:l})--(\ref{eq:n}). At the same time, $n_1$ has the smallest absolute value, corresponding to the fact that the nonlinear modulus $n$ depends only on the small difference of transverse sound velocities with polarization parallel and perpendicular to the applied pressure, according to Eq.~(\ref{eq:n}). The exponent $\gamma$ is found to lie in the range 2.4–2.9 for the polymers investigated in this study. It is noteworthy that, for relaxation systems characterized by a distribution of relaxation times, the linear response function cannot exhibit exponents $\gamma > 2$ \cite{Kremer-broadband-dielectric-spectroscopy-2003}. In contrast to the linear regime, in the case of nonlinear moduli, different relaxation processes can mutually interact, thereby generating arbitrary and more complex functional forms of the frequency dependence.

\color{black}

%\begin{align}
%    l(\omega) &= l_\infty + \frac{l_1}{1 + \omega^2 \tau^2},   \label{eq:lw} \\
%    m(\omega) &= m_\infty + \frac{m_1}{1 + \omega^2 \tau^2},   \label{eq:mw}  \\
%    n(\omega) &= n_\infty + \frac{n_1}{1 + \omega^2 \tau^2},   \label{eq:nw} 
%\end{align}
% \textcolor{red}{where $l_\infty$, $m_\infty$, and $n_\infty$ are high-frequency nonlinear moduli, $l_1$, $m_1$, and $n_1$ are contributions to the corresponding nonlinear moduli caused by the relaxation process with the characteristic time $\tau$}.
% Global fitting of the data obtained allowed for estimation of $\gamma$ for each of the three polymers. The obtained values are listed in Table \ref{tablefitting}.

%\begin{table}[]
%\centering
%\caption{Fitting parameters calculated for the three polymer materials assuming generalized Maxwell model with a single relaxation process.} 
%\begin{tabular}{|l|l|l|l|l}
%\cline{1-4}
%                   & PC        & PS        & PMMA      &  \\ \cline{1-4}
%$\tau$ ($\mu$s)    & 2712      & 3547      & 2508      &  \\ \cline{1-4}
%$n_{\infty}$ (Pa)  & -8.71E+09 & -6.89E+09 & -2.82E+10 &  \\ \cline{1-4}
%$n_1$    (Pa)      & 8.56E+15  & 1.62E+15  & 7.60E+16  &  \\ \cline{1-4}
%$m_{\infty}$  (Pa) & -8.09E+09 & -1.13E+10 & -2.82E+10 &  \\ \cline{1-4}
%$m_1$ (Pa)         & -4.77E+16 & -2.73E+16 & -4.54E+16 &  \\ \cline{1-4}
%$l_{\infty}$ (Pa)  & -2.76E+10 & -1.78E+10 & -6.28E+10 &  \\ \cline{1-4}
%$l_1$ (Pa)         & -1.31E+17 & -3.59E+17 & -3.45E+17 &  \\ \cline{1-4}
%\end{tabular}
%\label{tablefitting}
%\end{table}

\begin{table}[]
\caption{Fitting parameters of frequency dependencies of nonlinear elastic moduli of PC, PS and PMMA.}
\label{tablefitting}
% \begin{tabular}{|l|c|c|c|c|c|c|c|}
% \hline
%      & $l_{\infty}$(GPa) & $l_1$(GPa)& $m_{\infty}$(GPa)& $m_1$(GPa)& $n_{\infty}$(GPa)& $n_1$(GPa)& $\gamma$ \\ \hline
% PS   & -26        & -17   & -11        & -1.9  & -7.5       & 0.6   & -2.4     \\ \hline
% PMMA & -92        & -10   & -31        & -2.1  & -24        & 3.3   & -2.9     \\ \hline
% PC   & -33        & -10   & -9.2       & -3.8  & -8.1       & 0.7   & -2.6     \\ \hline
% \end{tabular}
\begin{tabular}{lccccccc}
\hline
         & $l_\infty$ & $l_1$ & $m_\infty$ & $m_1$ & $n_\infty$ & $n_1$ & $\gamma$ \\
         & GPa    & GPa·MHz$^\gamma\!$    & GPa      & GPa·MHz$^\gamma\!$ & GPa & GPa·MHz$^\gamma\!$ & \\ \hline
PC       & $-33$  & $-10$  & $-9.2$   & $-3.8$  & $-8.1$  & 0.7   & $2.6$ \\ 
PS       & $-26$  & $-17$  & $-11$    & $-1.9$  & $-7.5$  & 0.6   & $2.4$ \\
PMMA     & $-92$  & $-10$  & $-31$    & $-2.1$  & $-24$   & 3.3   & $2.9$ \\
\hline
\end{tabular}
\end{table}

\section{Conclusions}

Therefore, we performed the comparative experimental analysis of  frequency dependencies of linear (Lam\'e) and nonlinear (Murnaghan) elastic moduli of the three thermoplastic polymers, polystyrene, PMMA and polycarbonate, in the frequency range of 0.45--3 MHz. In all the three polymers the Lam\'e moduli demonstrated moderate rise with frequency; the most pronounced rise of about 35\% was observed in modulus $\lambda$ of PMMA.  The frequency dependencies of Murnaghan moduli were considerably nonlinear. At higher frequencies above $\sim$1 MHz no significant variations of the Murnaghan moduli occurred, while at lower frequencies the absolute values of the moduli $l$ and $m$ demonstrated rapid rise, more pronounced for the modulus $l$. At the same time that of the modulus $n$ decreased and demonstrated a tendency to become positive at lower frequencies. Note that both linear and nonlinear moduli of PMMA had higher values than those of PC and PS, with the latter two demonstrating close values of both types of moduli. 

The dominant factor governing the frequency dependencies of the moduli is believed to be the presence of relaxation processes and their sensitivity to the applied pressure. The observed dependencies indicate that these relaxation processes are more pronounced in the sub-MHz range. The global fit of the experimental data provided data on the values of the parameter $\gamma$, characterizing the dominating relaxation process.

\section{Declaration of competing interests}
The authors declare that they have no known competing financial interests or personal relationships that could have appeared to influence the work reported in this paper.

\section{Acknowledgements}
The financial support from Russian Science Foundation under the grant \# 22-72-10083-$\Pi$ is gratefully acknowledged.

%\appendix

%\section{Appendix}
%\label{app}
%Nonlinear terms of the one-dimensional equation of motion:

\bibliographystyle{model1-num-names}  
%\bibliographystyle{cas-model2-names}

\bibliography{cas-refs.bib}


\end{document}